\documentclass[aps,pre,twocolumn,showpacs,epsfig,epsf,amssymb,showkeys,tightenlines,floatfix]{revtex4-1}
\usepackage{graphicx}
\usepackage{amssymb}
\usepackage{amsmath}
\usepackage{subfigure}
\begin{document}

\title{ Effect of shearing on Wormlike micelles 
}

\author{Shaikh Mubeena}
\email{mubeena@students.iiserpune.ac.in}
\affiliation{
IISER-Pune, 900 NCL Innovation Park, Dr. Homi Bhaba Road,  Pune-411008, India.\\
}
\date{\today}
\begin{abstract}
   We use a hybrid method that combines the Multiparticle collision dynamics (MPCD) for solvent particles with the molecular dynamics for equilibrium polymers to simulate the shearing of the equilibrium polymers (or Wormlike micelles) at a mesoscopic length scale. The MPCD method incorporates the hydrodynamic interaction with the polymeric chains. We show successful implementation of the method on the model equilibrium polymers (or Wormlike micelles) and observe that the order of the Iso-Nem transition of the polymeric system is affected by the shear rate. Moreover, the chains of the equilibrium polymers first increase in their average length with the increase in shear rate but then show a decrease in their average length after crossing a particular value of the shear rate which shows the breaking of chains due to shear stress when their nematic order remains unchanged. This model and method can be further used to investigate the shear banding in Wormlike micelles or other interesting properties of such systems.

\end{abstract}
\keywords{soft matter, flow, hydrodynamic simulations}
\pacs{81.16.Dn,82.70.-y,81.16.Rf,83.80.Qr}
\maketitle
\section{Introduction}

     The equilibrium, non-equilibrium and dynamic behaviour of flexible, semi-flexible or rod-like polymers are strongly affected by the flow fields. They show large deformations and strong alignment under shear flow conditions ~\cite{smith1999single,gerashchenko2006statistics,winkler2006semiflexible,ripoll2006star,schroeder2005dynamics,huang2010semidilute}. The case of rod-like polymers is easier to study experimentally because of their large anisotropy as well as theoretically by including only translational and rotational motions ~\cite{hoffmann20093d}. However, including the hydrodynamic interactions is not always possible in theoretical studies. This field lacks a systematic study of the effect of hydrodynamic interactions on rods in dilute solutions in shear flow. 

    Including the conformational properties of flexible or semi-flexible polymers in analytical studies adds to the complexity of the analysis. Moreover, the inadequate inclusion of hydrodynamic properties has been a long-standing problem in analytical studies. Various techniques have been used to study the orientation and deformation of such molecules e.g. light scattering and neutron scattering, flow birefringence and non-equilibrium dynamics simulations ~\cite{fuller1995optical,janeschitz2012polymer,lang2016connection,aust2002rotation,harmandaris2014quantitative,ermak1978brownian,groot1997dissipative,kapral2008multiparticle}. The molecular dynamics simulations are able to implement hydrodynamic interactions successfully. However, the unnecessary evaluation of the solvent dynamics on colloidal or polymer length scales is a huge cost in time. Therefore, to bridge the gap between the large length and time scales, a coarse-grained simplified description of the solvent dynamics is used. e.g. lattice-gas, lattice-Boltzmann or multiparticle collision dynamics (MPCD) ~\cite{chen1998lattice,higuera1989boltzmann,kapral2008multiparticle}. 

     The multiparticle collision dynamics has been successfully implemented to study the non-equilibrium properties of colloids ~\cite{padding2004hydrodynamic,wysocki2009direct}, polymers ~\cite{ripoll2006star,winkler2004rod,ryder2006shear,chelakkot2010migration}, cells ~\cite{noguchi2005shape,mcwhirter2009flow} and vesicles ~\cite{noguchi2004fluid} in flow fields. In this paper, we study the effect of shear flow on equilibrium polymers (or Wormlike micelles) by using a hybrid simulation approach that combines the Molecular Dynamics (MD) simulations for polymers with the MPCD technique for solvent particles. We show that for the equilibrium polymeric system that undergoes an Isotropic-Nematic transition with an increase in its density, the effect of shear is to change the order of transition. The change in the Iso-Nem transition is abrupt for lower values of shear rates but the steepness of the slope at the transition decreases as the shear rate increases. Moreover, it is also shown that the polymeric chains first show an increase in their chain length and nematic order parameter with an increase in shear rate but then the chains start breaking after a particular value of shear rate when no further change in alignment is noted. Thus, we show a successful implementation of the hybrid method on the system of Wormlike micelles (or equilibrium polymers) undergoing shear flow. This method can be further used to investigate the interesting properties of the Wormlike micellar systems viz. shear thinning, shear thickening or shear banding.

\section{Model and method}

\subsection{modelling equilibrium polymers}

             We use a coarse-grained model of equilibrium polymers (or Wormlike micelles) that has been used in our earlier studies ~\cite{mubeena2015hierarchical,arxive1,arxive2,arxive3}. In this model, the following three potentials are used to form the Wormlike micelles or equilibrium polymers.\\

(1) $V_2$: Two body attractive L-J potential modified by an exponential term.\\
\begin{equation}      
        V_2 = \epsilon [ (\frac{\sigma}{r_2})^{12} - (\frac{\sigma}{r_2})^6 + \epsilon_1 e^{-a r_2/\sigma}]; 
\, \forall  r_2 < r_c.
\label{eq1}
\end{equation}

where, $r_2$ is the distance between monomers. The exponential term creates a potential barrier at $r_2=1.75\sigma$ for breaking (joining) the monomers from (to) a chain. We keep $\epsilon=110k_BT$, the cutoff distance $r_c=2.5\sigma$, $\epsilon_1=1.34\epsilon$ and $a= 1.72$. \\

(2) $V_3$: Three body potential to model semiflexibility of chains.\\
 For any 3 monomers with a central monomer bonded with two other monomers at a distance $r_2$ and $r_3$ and forming an angle $theta$ at the central monomer, the following 3body potential adds the semiflexibility to the chains, 

\begin{equation}
V_3 = \epsilon_3 (1 - \frac{r_{2}}{\sigma_3})^2(1 - \frac{r_{3}}{\sigma_3})^2 \sin^2(\theta); 
\, \forall r_{2},r_{3} < \sigma_3. 
\label{eq2}
\end{equation}

where, $\epsilon_{3}=6075k_{B}T$, and the cutoff distance $\sigma_3=1.5\sigma$. \\

(3) $V_4$: Four body potential to avoid branching.\\

    To avoid branching, a four body potential is used to repel any chain trying to form a branch,

\begin{equation}
V_4 = \epsilon_4 (1 - \frac{r_{2}}{\sigma_3})^2(1 - \frac{r_{3}}{\sigma_3})^2 \times V_{LJ}(\sigma_4,r_4) 
\label{eq3}
\end{equation}

where, $r_2$ and $r_3$ are the distances of the two bonded monomers from the central monomer in a chain and $r_4$ is the distance from the monomer attached to the other chain that needs to be repelled. This is a shifted L-J potential with only the positive part. The cutoff distance for this potential $\sigma_4$ is chosen such that $ \sigma_3 < \sigma_4 < r_c$ and is fixed at $\sigma_4=1.75\sigma$. This model has been already used to show the Iso-Nem transition and the exponential length distribution of the polymeric chains confirming the characteristic properties of the Wormlike micellar/equilibrium polymeric system. Please refer ~\cite{mubeena2015hierarchical,arxive1} for a detailed description of the model and its successful implementation.

\section{Method:}

We use Molecular dynamic (MD) method to evolve equilibrium polymers and is coupled with the MPCD technique to simulate the hydrodynamic effect.
The solvent is composed of $N_s$ point particles of mass $m_s$ interacting with each other by a stochastic process. The MPCD technique to simulate solvent particles consists of two steps: streaming step and collision step. 
In the streaming step, the position of a particle i at any time t, with a velocity $\bar{v_i(t)}$ is updated to a time t+h according to,

\begin{equation}
\bar{r_i}(t+h) = \bar{r_i}(t) + h\bar{v_i}(t);
\,  \forall i=1,...,N_s.
\end{equation}

In the collision step, the simulation box is divided into cubic cells of length a and the particles are sorted into these cells. The centre of mass velocity in each cell is calculated and the relative velocities of particles with respect to the centre of mass velocity of the cell are rotated around a randomly oriented axis by an angle of $130^{\circ}$.

\begin{equation}
	\bar{v_i}(t+h) = \bar{v_i}(t) + (\bar{R}(\alpha)-\bar{E})(\bar{v_i}(t)-\bar{v}_{cm}(t))
\end{equation}

where, $\bar{R}(\alpha)$, $\bar{E}$ and $\bar{v}_{cm}$ are the rotation matrix, the unit matrix and the centre of mass velocity of the cell in which the particle is present. Here, $\alpha=130^\circ$.

The effect of hydrodynamic on the polymers is taken into account by including polymers in the collision step. Therefore, in a collision cell c with ${N_c}^m$ monomers of mass $m_m$ and ${N_c}^s$ solvent particles of mass $m_s$, the centre of mass velocity is calculated to be,

\begin{equation}
v_{cm}(t) = \frac{{\sum_{i=1}}^{{N_c}^s} m_s\bar{v_i}(t) + {\sum_{j=1}}^{{N_c}^m}m_m\bar{v_j}(t)}{{m_s{N_c}^s}+{m_m{N_c}^m}}
\end{equation}

At every collision step, a random shift is performed to ensure the Galilean invariance. It is ensured that the mass and momentum is conserved in the collision step.
The shear flow is imposed using Lees-Edwards boundary conditions along the y-direction. The shear flow is applied by two oppositely moving planes aligned with the x-z plane. A no-slip boundary or a bounce back condition ensures that the veloicty of any particle (monomer or solvent) colliding with the wall is reverted. 

The fluid temperature is maintained by using a local Maxwellian thermostat. We fix $k_BT=1$.
With the parameters $k_BT=1$, collision time $h=0.1$, $m_s=1$ and collision cell size $a=1$, the shear viscosity yields the value of 8.7. 

The equilibrium polymers are evolved using Molecular Dynamics(MD) simulations. For each collision step h, the molecular dynamics to evolve polymers is called for $h/h_p$ steps with $h_p=0.002$.

The diameter of monomers $\sigma=1$ is chosen as the unit of length in this paper. The simulation box is a cubic one with dimensions $30\times30\times 30\sigma^{3}$. The monomer number density is denoted by $\rho_m\sigma^{-3}$ which depicts the number of monomers in a simulation box.
The solvent particle density is kept to be 10 particles per unit collision cell with the mass of the particle $m_s=1$.

The length and time are scaled according to $\hat{r}=r/a$ and $\hat{t}=t\sqrt{k_BT/m_sa^2}$.

The monomer number density and the shear rate is varied and the corresponding changes are observed which are presented in the next section.

\section{Results ::}

   The method has been first checked by applying walls along x-z plane and providing a force in the x-direction. Rejecting first few thousand runs and then averaging over next $10^5$ runs, a velocity profile is generated. The component of velocities along the x-axis are plotted against y-direction and it generated a Poiseuille profile. The velocity profile has a small wall slip with just a bounce-back condition. When phantom particles are taken into account in the wall, the wall slip is removed.

\begin{figure}
\centering
\includegraphics[scale=0.3]{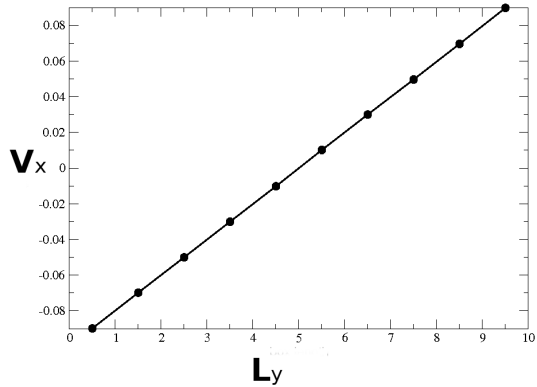}
\caption{ The figure shows the velocity profile generated by applying Lees edwards boundary condition in the shearing direction for a shear rate of $0.02\tau^{-1}$ in a box of size $10\times 10\times 10\sigma^3$ with MPCD fluid particles number density of $10\sigma^{-3}$.}
\label{shear_prof}
\end{figure}

Now the walls are replaced with the Lees-Edwards boundary conditions to generate a shear flow. The velocity profile for the shear flow is shown in Fig. \ref{shear_prof} for a shear rate of $0.02\tau^{-1}$ in a $10\times 10\times 10 \sigma^3$ simulation box. 

\begin{figure}
\centering
\includegraphics[scale=0.2]{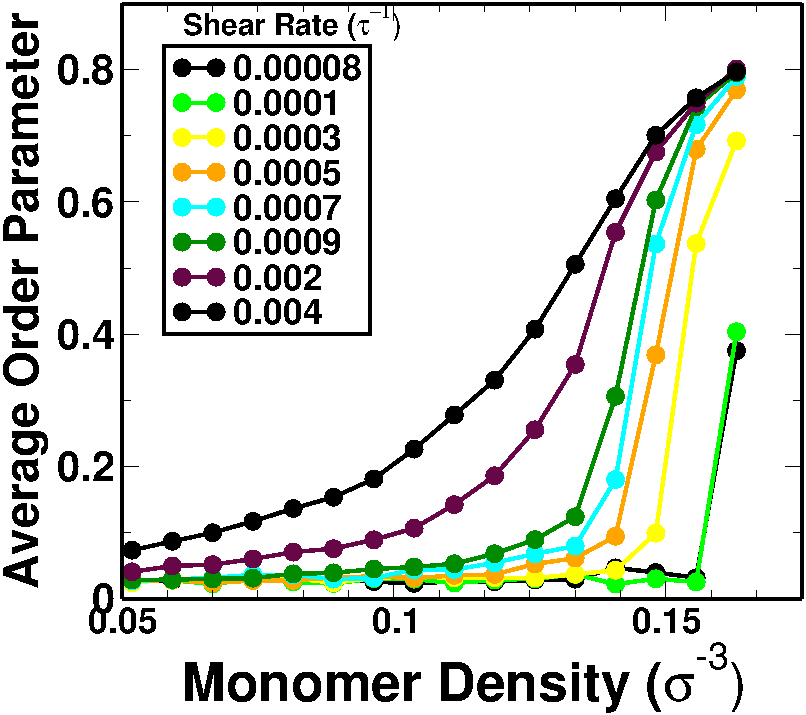}
\includegraphics[scale=0.2]{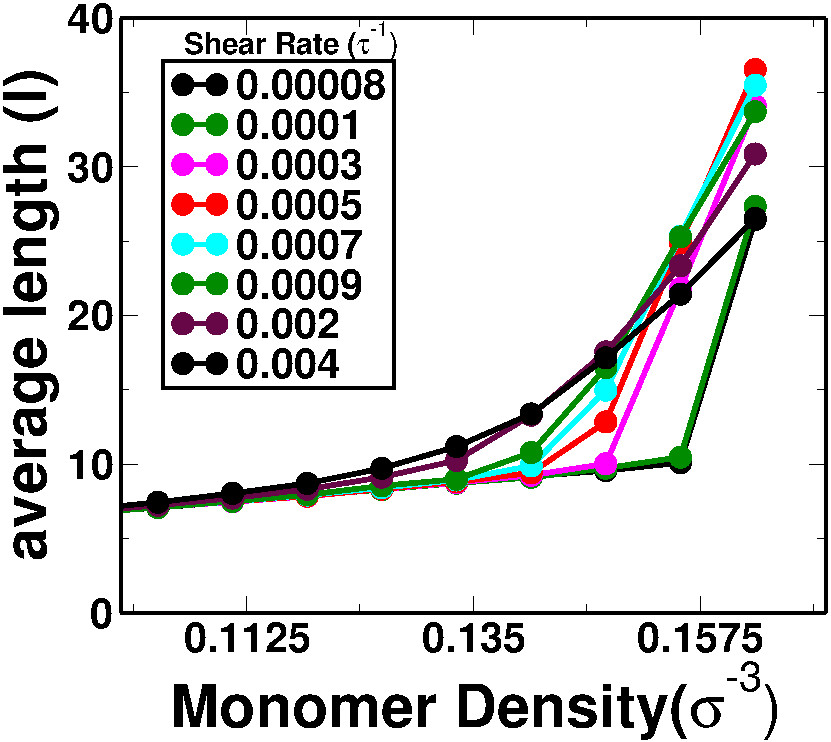}
\caption{ The figure shows the graphs of (a) average nematic order parameter and (b) average length of micellar chains plotted against the micellar densities for diferent values of shear rates as indicated by different symbols in the figures. Both the quantites show a slow increase in their values for lower densities of micelles, but a sudden change in their values at a higher density indicate an isotropic to nematic transition. The transition point gets shifted to lower value of density with increase in the shear rate. Moreover, the steepness of the slope at the transition is higher for lower shear rates.}
\label{obs}
\end{figure}

Thus, the Poiseuille profile and the shear profile confirms the method used. Now, we include the monomers in the simulation box and using the hybrid method (coupling molecular dynamics for monomers with multiparticle collision dynamics) shear the equilibrium polymers in an MPCD fluid. The shear rate is varied from $0.00008\tau^{-1}$ to $0.002\tau^{-1}$. For each value of shear rate, the monomer number density $\rho_m$ is also varied and the changes in the average nematic order parameter $<S>$ and the average length of the polymeric chains $<L>$ are calculated. The average nematic order parameter is the measure of the alignment of polymeric chains and is calculated by taking the average of $cos^2\theta$ where $\theta$ is the angle that polymeric chains make with the aligning axis.

\begin{figure}
\centering
\includegraphics[scale=0.2]{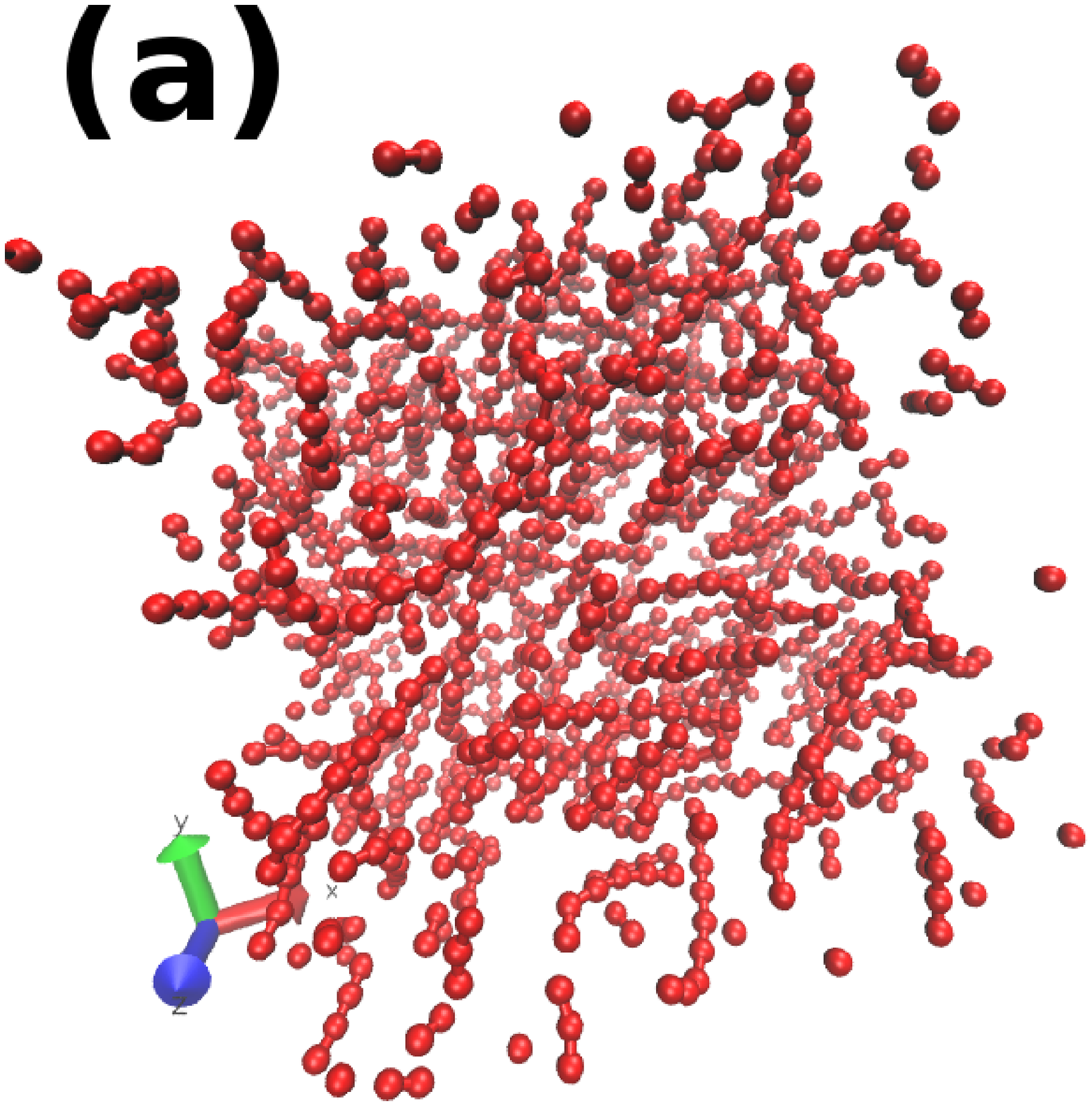}
\includegraphics[scale=0.2]{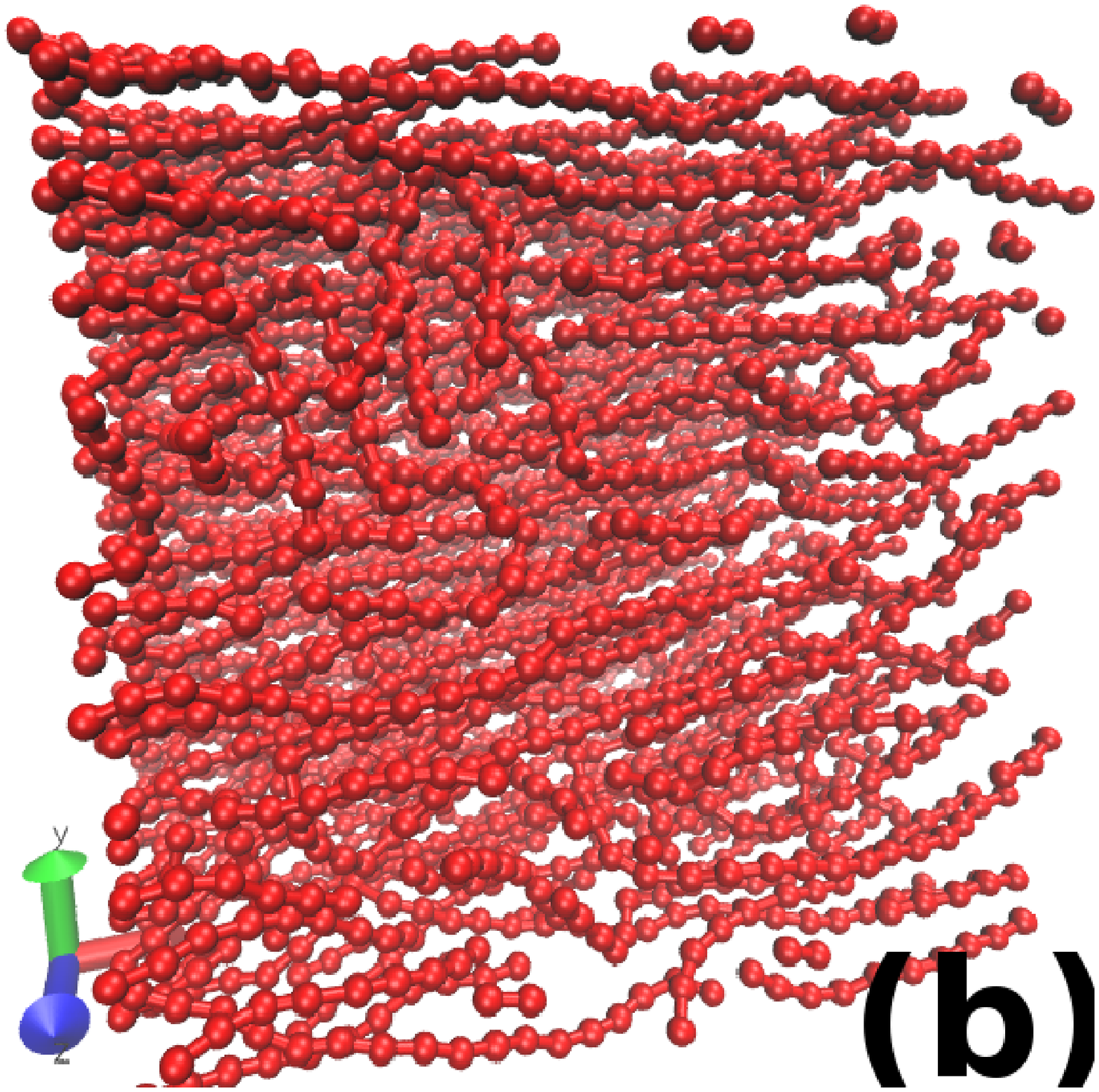}
\caption{ The figure shows the snapshots of model Wormlike micelles in a $30\times 30\times 30\sigma^3$ box subjected to the shear with a rate of $0.002\tau^{-1}$ with two different monomer densities in the box (a) $\rho_m=0.059\sigma^{-3}$ and (b) $\rho_m=0.148\sigma^{-3}$. With increase in the density of micelles the micellar nematic order parameter is seen to be increasing and producing an aligned phase of micelles at a higher density. Moreover, with increase in the density the average length of micellar chains also increases. }
\label{snap1}
\end{figure}

\begin{figure}
\centering
\includegraphics[scale=0.2]{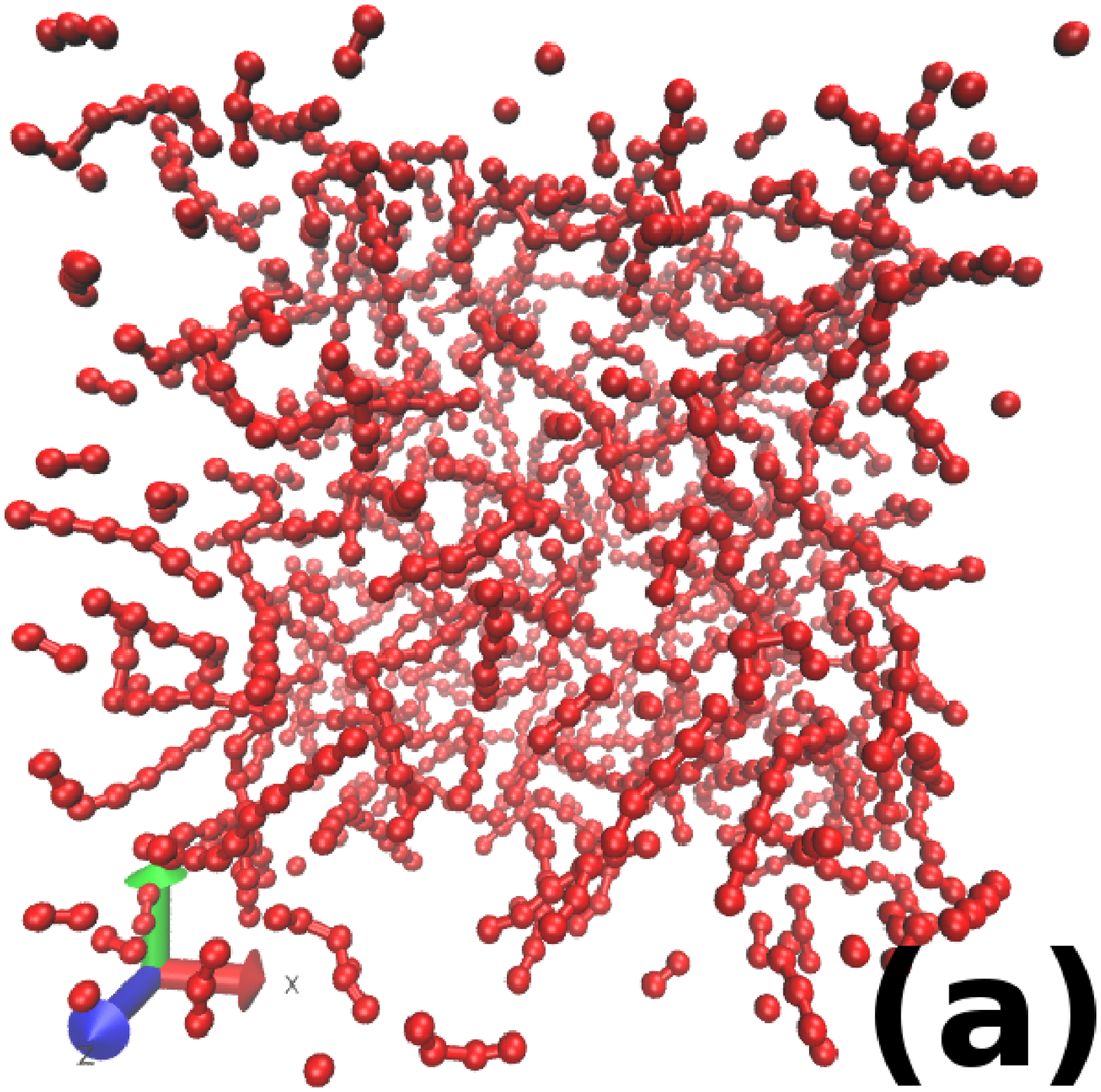}
\includegraphics[scale=0.2]{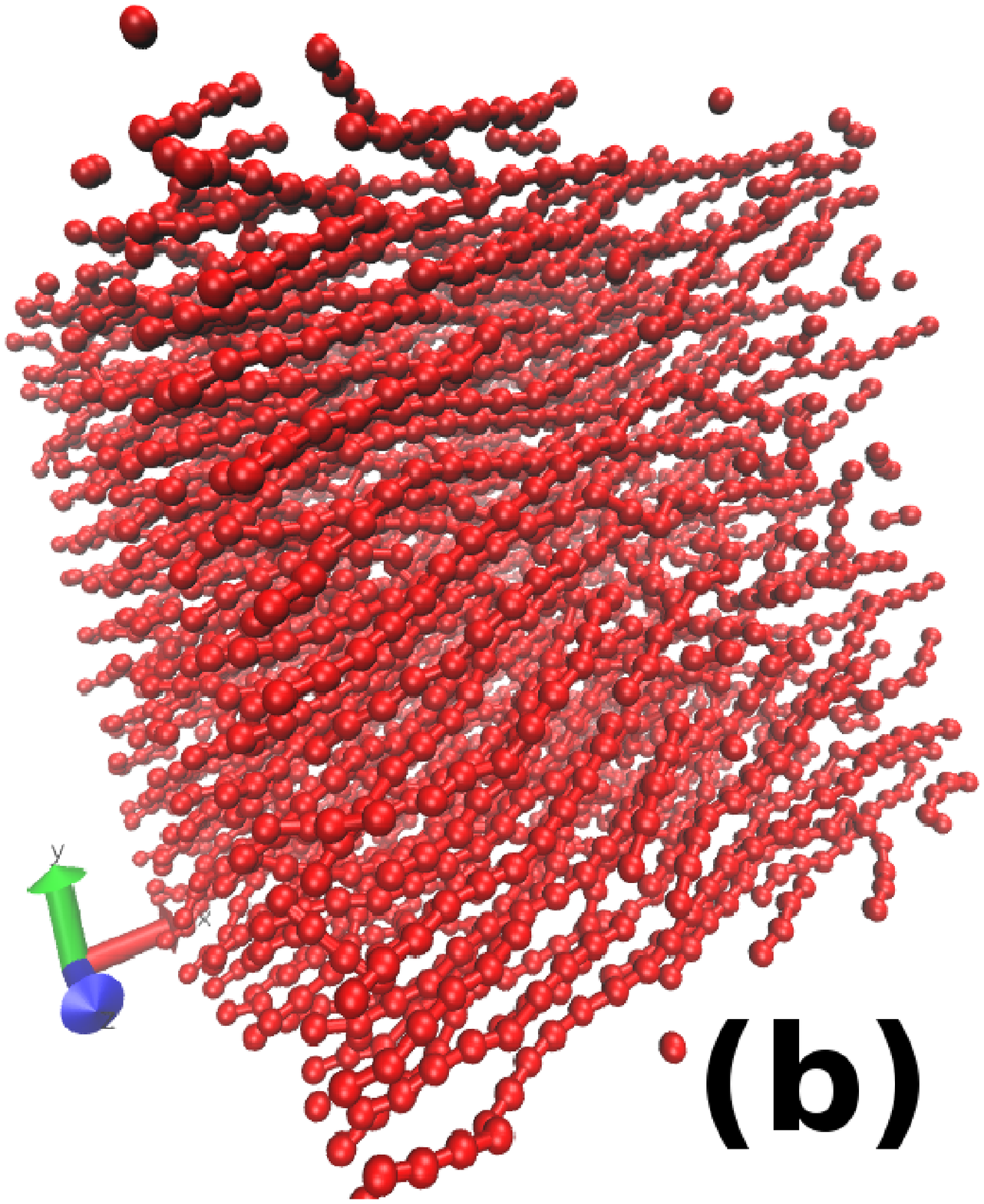}
\caption{ The figure shows two snapshots of Wormlike micelles subjected to the shear rate of $0.0009\tau{-1}$ with monomer number densities (a) $0.059\sigma^{-3}$ and (b) $0.148\sigma^{-3}$. With increase in the value of monomer density, the average length and the nematic order parameter of micellar chains increases as shown in the graphs of Fig.\ref{obs}}
\label{snap2}
\end{figure}

 These results are shown in Fig. \ref{obs}(a) and (b). The nematic order parameter in Fig.\ref{obs}(a) shows an abrupt change in its value as the monomer number density is increased, only for low values of shear rates: shear rate=0.00008$\tau^{-1}$ and shear rate=0.0001$\tau^{-1}$. These abrupt changes correspond to the Isotropic-Nematic transitions. As the shear rate increases further, the transition becomes more and more continuous. Thus, the effect of shear rate here is to change the order of the transition. Similar observations are shown in Fig.\ref{obs}(b) where the average length of the polymeric chains is shown. Moreover, the transition points get shifted to lower values of $\rho_m$ as the shear rate increases.

These results are confirmed by observing the snapshots of the polymers. Two of the snapshots are shown in Fig.\ref{snap1}.The red particles indicate the monomers. The fig.\ref{snap1}  shows the snapshots for two different values of $\rho_m$: (a) $0.059\sigma^{-3}$ and (b) $0.148\sigma^{-3}$ with the shear rate fixed at $0.002\tau^{-1}$. It can be clearly seen that the polymeric chains are smaller and in the isotropic state for lower monomer density in (a) while the polymeric chains in figure(b) have a higher chain length and aligned with each other.
Figure \ref{snap2} shows one more set of snapshots for the same densities considered in fig.\ref{snap1} but for shear rate$=0.0009\tau^{-1}$.
Comparing the Figs.\ref{snap1} and \ref{snap2}, we observe that snapshots with same monomer densities show longer chains and higher nematic ordering for lower shear rates.

\begin{figure}
\centering
\includegraphics[scale=0.2]{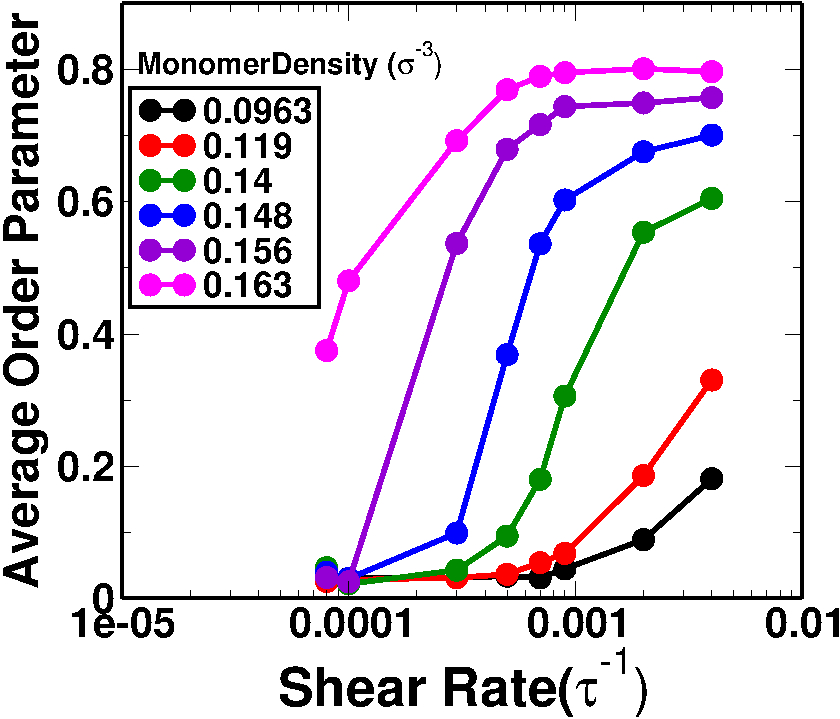}
\includegraphics[scale=0.2]{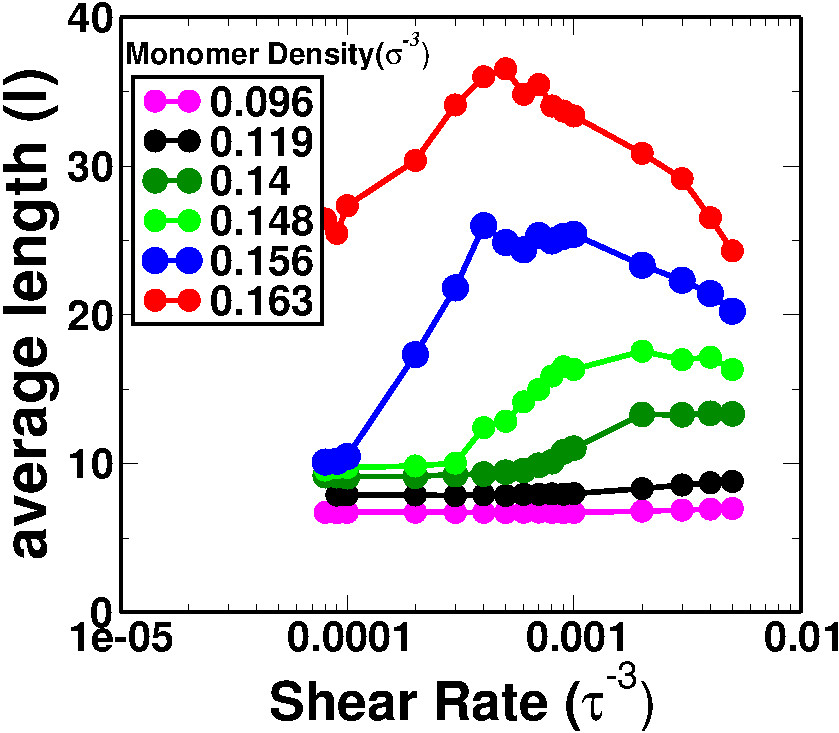}
\caption{ The figure shows two plots for average nematic order parameter $<S>$ in (a) and average length $<L>$ in (b) of micellar chains plotted againt the shear rates. Each figure shows different graphs for different values of monomer number density $\rho_m$. For each case of micellar density, the average order parameter first show an increase in its value and then remains constant after certain value of shear rate. The graphs for average length of monomers first show an increase with increase in shear rate, but decrease after reaching a certain value of shear rate  after which the average order parameter of micelles show a constant value. }
\label{sh_rate}
\end{figure}

 It is also noted for other values of $\rho_m$ that the average length increases with an increase in the shear rate but then decreases with further increase in shear rate beyond the isotropic-nematic transition point. This can be observed directly in Fig.\ref{obs} but easier to observe by plotting these parameters v/s shear rates. This is shown in Fig.\ref{sh_rate} (a) $<S>$ and (b) $<L>$ plotted against shear rates for different values of monomer number densities $\rho_m$. For each value of $\rho_m$, the order parameter in figure(a) can be seen to be increasing with increase in shear rate and then shows a constant value when the polymers are well aligned with each other and $<S>$ gets saturated. It can be seen in (b) that the average length increases with an increase in shear rate but shows a decrease in its value with further increase in shear rate after a particular value of shear rate. This indicates that the polymeric chains join with each other and increases in their length with an increase in the shear rate for the lower value of shear rates. However, after a particular value of shear rate, the polymeric chains start breaking due to shear stress.

\bibliographystyle{apsrev4-1}
\bibliography{paper_shear_mono}

\end{document}